\documentclass[aps,prb,twocolumn,preprintnumbers,amsmath,amssymb,superscriptaddress]{revtex4}%

\usepackage{graphicx}%
\usepackage{dcolumn}
\usepackage{amsmath}
\usepackage{color}

\newcommand{\RN}[1]{\textup{\uppercase\expandafter{\romannumeral#1}}}%

\begin{document}

\title{On the superconducting nature of the Bi-II phase of elemental Bismuth}

\author{Rustem~Khasanov}
 \email{rustem.khasanov@psi.ch}
 \affiliation{Laboratory for Muon Spin Spectroscopy, Paul Scherrer Institute, CH-5232 Villigen PSI, Switzerland}

\author{Milo\v{s} M. Radonji\'{c}}
\affiliation{Scientific Computing Laboratory, Center for the Study of Complex Systems, Institute of Physics Belgrade, University of Belgrade, Pregrevica 118, 11080 Belgrade, Serbia}

\author{Hubertus Luetkens}
 \affiliation{Laboratory for Muon Spin Spectroscopy, Paul Scherrer Institute, CH-5232 Villigen PSI, Switzerland}

\author{Elvezio Morenzoni}
 \affiliation{Laboratory for Muon Spin Spectroscopy, Paul Scherrer Institute, CH-5232 Villigen PSI, Switzerland}

\author{Gediminas Simutis}
 \affiliation{Laboratory for Muon Spin Spectroscopy, Paul Scherrer Institute, CH-5232 Villigen PSI, Switzerland}

\author{Stephan Sch\"{o}necker}
    \affiliation{Applied Materials Physics, Department of Materials Science and Engineering, KTH Royal Institute of Technology, SE-10044 Stockholm, Sweden}

\author{Wilhelm H. Appelt}
    \affiliation{Augsburg Center for Innovative Technologies, and Center for Electronic Correlations and Magnetism, Theoretical Physics III, Institute of Physics, University of Augsburg, D-86135 Augsburg, Germany}

\author{Andreas \"{O}stlin}
    \affiliation{Augsburg Center for Innovative Technologies, and Center for Electronic Correlations and Magnetism, Theoretical Physics III, Institute of Physics, University of Augsburg, D-86135 Augsburg, Germany}

\author{Liviu Chioncel}
    \affiliation{Augsburg Center for Innovative Technologies, and Center for Electronic Correlations and Magnetism, Theoretical Physics III, Institute of Physics, University of Augsburg, D-86135 Augsburg, Germany}

\author{Alex Amato}
 \affiliation{Laboratory for Muon Spin Spectroscopy, Paul Scherrer Institute, CH-5232 Villigen PSI, Switzerland}

\begin{abstract}
The superconductivity in the Bi-II phase of elemental Bismuth (transition temperature $T_{\rm c}\simeq3.92$~K at pressure $p\simeq 2.80$~GPa) was studied experimentally by means of the muon-spin rotation as well as theoretically by using the Eliashberg theory in combination with
Density Functional Theory calculations. Experiments reveal that Bi-II is a type-I superconductor with a zero temperature value of the thermodynamic critical field $B_{\rm c}(0)\simeq31.97$~mT. The Eliashberg theory approach provides a good agreement with the experimental $T_{\rm c}$
and the temperature evolution of $B_{\rm c}$.  The estimated value for the retardation (coupling) parameter $k_{\rm B}T_{\rm c}/\omega_{\rm ln} \approx 0.07$ ($\omega_{\rm ln}$ is the logarithmically averaged phonon frequency) suggests that Bi-II is an intermediately-coupled superconductor.
\end{abstract}


\maketitle

Bismuth is element 83 in the periodic table.  It is a brittle metal with a silvery white color.
Its complex and tunable electronic structure exhibits many fascinating properties that often defy the expectations of conventional theories of metals.
Most notably, measurements on Bismuth provided the first evidence of quantum oscillations and the existence of the Fermi surface, thereby experimentally confirming the underlying paradigm of all modern solid state physics.\cite{Shubnikov_Nature_1930,deHaas_PNRAS_1930}

At ambient pressure Bismuth is a compensated semimetal with an exceptionally low carrier concentration of  one free charge carrier
per about $10^5$ atoms.\cite{Yang_PRB_2000} The Fermi surface consists of tiny electron- and hole-like pockets giving rise to a highly-anisotropic effective mass, which can become as low as $\sim10^{-3}$ that of the electron mass in some directions.\cite{Hofmann_PiSS_2006} Such properties lead to the highest Hall coefficient, the largest diamagnetism, and an exceptionally small thermal conductivity which sets Bismuth to be quite different compared to other metals.\cite{Norman_book_1998}

Upon application of pressure at room temperature, Bi undergoes a series of structural transitions:\cite{Degtyareva_HPR_2005}
\begin{equation}
{\text{Bi-I}} \stackrel{{\rm 2.55~GPa}}{\longrightarrow}
{\text{Bi-II}} \stackrel{{\rm 2.7~GPa}}{\longrightarrow}
{\text{Bi-III}} \stackrel{{\rm 7.7~GPa}}{\longrightarrow}
{\text{Bi-V}}< {\rm 220~GPa}. \nonumber
\end{equation}
Upon cooling, all the above phases become superconducting with the transition temperature ($T_{\rm c}$) of $T_{\rm c}\simeq0.53$~mK for Bi-I, $T_{\rm c}\simeq3.9$~K for Bi-II, $T_{\rm c}\simeq 7$~K for Bi-III and $T_{\rm c}\simeq 8.5$~K for Bi-V,
respectively.\cite{Brandt_JETP_1963, Buckel_PL_1965, Ilina_JETP_1970, Lotter_EPL_1988, Du_PRL_2005, Prakash_Science_2017, Li_PRB_2017, Brown_ScAdv_2018, Brown_Thesis_2017, Khasanov_PRB-Bi-III_2018}
The superconductivity in Bi-I and Bi-III phases were found to be of a type-I and type-II, respectively.\cite{Prakash_Science_2017,Li_PRB_2017, Brown_ScAdv_2018, Brown_Thesis_2017, Khasanov_PRB-Bi-III_2018}  Much less information is known for other Bi phases. In particular, the Bi-I to Bi-II and Bi-II to Bi-III  transitions are well-established at room-temperature, while their low temperature behavior lead to contradicting results. References~\onlinecite{Bundy_PR_1958, Klement_PR_1963, Il'ina_PSS_1967, Compy_JAP_1970, Yomo_JPSJ_1972, Homan_JPCS_1975} suggest that the Bi-III phase forms at 2.7~GPa at room temperature, while the Bi-II to Bi-III (or possibly Bi-I to Bi-III) phase boundary occurs  at pressures $p\gtrsim 3.0$~GPa at 0~K. The Bi-II phase likely extends down to 200~K only, where the Bi-I-II-III triple point may occur.\cite{Compy_JAP_1970, Homan_JPCS_1975} On the other hand, the superconducting Bi-III phase was observed at pressures of $\simeq$2.7~GPa by several other research groups, as well as by us.\cite{Li_PRB_2017, Brown_ScAdv_2018, Brown_Thesis_2017, Khasanov_PRB-Bi-III_2018} Some groups have also reported
superconductivity in Bi-II phase at pressures of $\simeq 2.5$~GPa with $T_{\rm c}\simeq4$~K.\cite{Brandt_JETP_1963, Homan_JPCS_1975, Lotter_EPL_1988} It is worth to note here,  that a pure Bi-II phase has never been observed alone, but always appeared as an admixture to the Bi-I or Bi-III phases.\cite{Li_PRB_2017, Brown_Thesis_2017, Lotter_EPL_1988}  It seems therefore likely, that the Bi-II phase becomes metastable at low temperatures.

\begin{figure*}[tb]
\hspace{-10mm}
\includegraphics[width=1.05\linewidth]{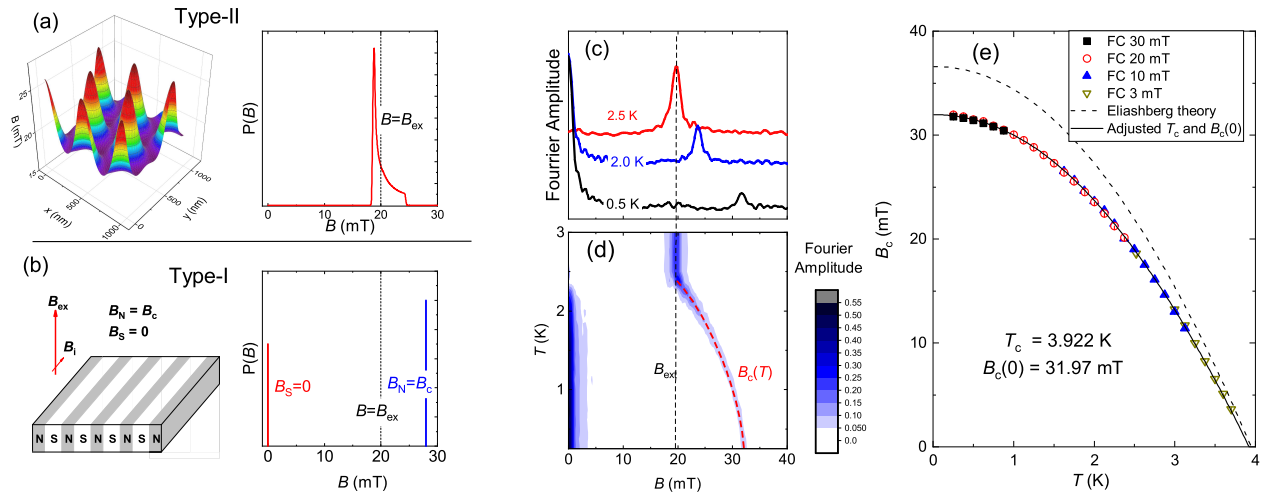}
\vspace{-1.3cm}
\caption{(a) The distribution of fields in a type-II superconductor in the  vortex state (left panel) and the corresponding magnetic field distribution function $P(B)$ (right panel). (b) The schematic representation of nucleation of a plate-like type-I superconductor in normal state ($B_{\rm N}=B_{\rm c}$) and superconducting ($B_{\rm S}=0$) domains (left panel). An ordered laminar structure is formed with an additional in-plane component $B_i$, after Refs.~\onlinecite{Sharvin_JETP_1958, Dorsey_PRB_1998}. The right panel is the $P(B)$ distribution in type-I superconductor. (c) Fourier transform of TF-$\mu$SR time spectra measured at external field $B_{\rm ex}\simeq 20$~mT reflecting the $P(B)$ distribution in Bi-II sample above ($T=2.5$~K) and below ($T=0.5$ and 2.0~K) the superconducting transition temperature [$T_{\rm c}({\rm 20~mT})\simeq 2.3$~K]. (d) The contour plot of the $P(B)$ distribution measured at $B_{\rm ex}\simeq 20$~mT. (e) The temperature dependence of the thermodynamical critical field $B_{\rm c}$ for Bi-II sample obtained in $\mu$SR experiments with the applied field $B_{\rm ex}=3$, 10, 20, and 30~mT. The dashed line is the temperature evolution of the 'theoretical' $B_{\rm c, T}$ obtained within the framework of {\it ab-initio} Eliashberg calculations using Density Functional Theory. The solid line is the same $B_{\rm c, T}(T)$ curve with $T_{\rm c}=3.922$~K and $B_{\rm c}(0)=31.97$~mT adjusted from the fit (see text for details).  }
 \label{fig:experiment}
\end{figure*}

This paper presents the results of an experimental and theoretical study of the Bi-II superconductor. The bulk Bi-II phase ($T_{\rm c}\simeq 3.92$~K at $p\simeq 2.80$~GPa) was stabilized by approaching it from the preformed Bi-III one ($T_{\rm c}\simeq 7.05$~K at $p\simeq 2.72$~GPa, Ref.~\onlinecite{Khasanov_PRB-Bi-III_2018}).  Muon-spin-rotation ($\mu$SR) measurement reveal that the magnetic induction ($B$) in a cylindrical Bi-II sample (with the magnetic field applied perpendicular to the cylinder axis) is separated between normal state ($B_{\rm N}=B_{\rm c}$, $B_{\rm c}$ is the thermodynamical critical field) and superconducting ($B_{\rm S}=0$) domains thus indicating that Bi-II is a  superconductor of type-I. The zero temperature thermodynamic critical field was found to be $B_{\rm c}(0)\simeq31.97$~mT. The Eliashberg theory provides a good agreement with the experimental critical temperature ($\simeq 3.95$~K), the zero temperature critical field ($\simeq 36.6$~mT), and the temperature evolution of $B_{\rm c}(T)$.  The estimated value for the retardation parameter $k_{\rm B}T_{\rm c}/\omega_{\rm ln} \approx 0.07$ ($\omega_{\rm ln}$ is the logarithmically averaged phonon frequency) suggests that Bi-II is an intermediately-coupled superconductor.

The Bi sample and the pressure cell were the same as used in our previous experiments for studying  Bi-III superconductivity, Ref.~\onlinecite{Khasanov_PRB-Bi-III_2018}. The transformation of the Bi sample from Bi-III to Bi-II phase was made by allowing the sample volume to increase inside the pressure cell.\cite{Supplemntal_part} 
ac susceptibility (ACS) measurements reveal the presence of a sharp superconducting transition at $T_{\rm c}\simeq 3.92$~K at $p\simeq 2.80$~GPa. The amount of Bi-III phase admixture, obtained in the ACS experiments, does not exceed $10-15$~\%  (see the Supplemental Part, Ref.~\onlinecite{Supplemntal_part}).
The transverse-field (TF) $\mu$SR experiments were  carried out at the $\mu$E1 beam line by using the dedicated GPD (General Purpose Decay) spectrometer (Paul Scherrer Institute, Switzerland). The details of   TF-$\mu$SR experiments performed under pressure are provided in the Supplemental Part, Ref.~\onlinecite{Supplemntal_part}, and in Refs.~\onlinecite{Khasanov_PRB_2011, Khasanov_HPR_2016, Shermadini_HPR_2017}.

\begin{table*}[htb]
\caption{\label{tab1} Experimental and calculated material parameters for various Bismuth phases. $T_{\rm c}$ is the superconducting transition temperature,  $B_{\rm c}$ is the thermodynamical critical field, $B_{\rm c2}$ is the upper critical field,  $\lambda_{\rm el-ph}$ is the electron-phonon coupling constant, $\gamma_{\rm N}$ is the normal state electronic specific heat coefficient,\cite{Supplemntal_part} $\omega_{\rm ln}$ is the characteristic phonon frequency, and $k_{\rm B} T_{\rm c}/\omega_{\ln}$ is the retardation (coupling) parameter. n/a means the parameter is not available.    }
\begin{tabular}{lcccccccccc}
\hline
\hline
&Superconductivity&$T_{\rm c}$&$B_{\rm c}$&$B_{\rm c2}$&$\lambda_{\rm el-ph}$&$\gamma_{\rm N}$&$\omega_{\rm ln}$&$k_{\rm B} T_{\rm c}/\omega_{\ln}$& References\\
&                 &(K)                          & (mT)      & (T)        & &(${\rm erg}\;{\rm cm}^{-3}{\rm K}^{-2}$)&&(meV)&&\\
\hline
Bi-I   & type-I    & 0.00053                         &0.0052     & --         &0.236& 399&n/a&n/a&\onlinecite{Prakash_Science_2017, Collan_PRL_1969, Mata-Pizon_Plos_2016}\\
Bi-II  & type-I    & 3.92                              &31.97      & --         &1.02& 2206&4.69& 0.072 &This Work\\
Bi-III & type-II   & 7.05                             &73.6       &2.6         &2.75&n/a &5.51& 0.110 &\onlinecite{Brown_ScAdv_2018, Brown_Thesis_2017, Khasanov_PRB-Bi-III_2018}\\
Bi-V   &   n/a       & 8.50                          &n/a          &n/a           &n/a&n/a&n/a&n/a&\onlinecite{Lotter_EPL_1988, Brown_Thesis_2017}\\
\hline
\hline
\end{tabular}
\end{table*}

Due to its microscopic nature, the $\mu$SR technique allows to directly distinguish between type-I and type-II superconductors, since both superconductivity types are characterized by very different magnetic field distributions [$P(B)$'s] inside the specimen. An ordered flux-line lattice (FLL) of type-II superconductor has the field distribution and the corresponding $P(B)$ which are shown schematically in the left and right panels of Fig.~\ref{fig:experiment}~(a). The calculations were performed within the framework of London model with the Gaussian cutoff for a triangular FLL ($B_{\rm ex}=20$~mT, the magnetic penetration depth $\lambda=200$~nm and the coherence length $\xi=50$~nm, see the Supplemental Part, Ref.~\onlinecite{Supplemntal_part}). The asymmetric magnetic field distribution function $P(B)$  centers in the vicinity of $B_{\rm ex}$. It is characterized by two cutoffs fields and by the peak shifted below $B_{\rm ex}$ [see the right panel at Fig.~\ref{fig:experiment}~(a), and, {\it e.g.}, Refs.~\onlinecite{Maisuradze_JPCM_2009, Khasanov_PRB_2016} and references therein].
A type-I superconductor expels a magnetic field completely, apart from a layer at the surface of thickness $\lambda$.
However in samples with a finite demagnetization factor $n$, a separation between superconducting domains  (with $B_{\rm S}=0$) and normal state domains (with $B_{\rm N}=B_{\rm c}>B_{\rm ex}$) can occur [see the left panel at Fig.~\ref{fig:experiment}~(b) showing schematically the nucleation of a plate-like sample on S/N domains, and, {\it e.g.}, Refs.~\onlinecite{Tinkham_75, Huebener_Book_1979, Prozorov_PRL_2007, Prozorov_NatPhys_2008} and references therein]. In this case, $P(B)$ consists of two, $B=0$ and $B=B_{\rm c}$, lines [right panel of Fig.~\ref{fig:experiment}~(b)]. Such distributions (without, however, the $B=0$ line)    were reported in earlier $\mu$SR measurements on type-I superconductors Sn, Pb and In,\cite{Gladisch_HypInt_1979, Grebinnik_JETP_1980, Egorov_PhysB_2000, Egorov_PRB_2001} and in recent experiments on BeAu.\cite{Singh_Arxiv_2019,Beare_Arxiv_2019}

Figure~\ref{fig:experiment}~(c) shows the Fourier transform of few representative TF-$\mu$SR time spectra (the pressure cell background subtracted) measured at $B_{\rm ex}=20$~mT. Figure~\ref{fig:experiment}~(d) represents the contour plot of the corresponding Fourier intensities. The overall behavior shown in Figs.~\ref{fig:experiment}~(c) and (d) corresponds to the response of a type-I superconductor with a nonzero demagnetization factor $n$ in an applied field $B_{\rm ex}$ of $B_{\rm c} (1-n)\leq B_{\rm ex} \leq B_{\rm c}$ [see the discussion above, Fig.~\ref{fig:experiment}~(b) and Refs.~\onlinecite{Gladisch_HypInt_1979, Grebinnik_JETP_1980, Egorov_PhysB_2000, Egorov_PRB_2001,Singh_Arxiv_2019,Beare_Arxiv_2019}]. Indeed, the $P(B)$ distributions at $T\simeq0.5$ and  2.0~K split into  two  peaks with the first one at $B=0$ and the second one $\simeq 12$ and $\simeq 5$~mT higher than the applied field $B_{\rm ex}$, respectively.  With increasing temperature, the intensity of the $B=0$ peak  decreases until it vanishes at $T\simeq 2.3$~K, while  the  intensity of the $B\geq B_{\rm ex}$ peak  increases by approaching $T\simeq2.3$~K and saturates above it [Figs.~\ref{fig:experiment}~(c) and (d)]. The position of the $B\geq B_{\rm ex}$ peak shifts in direction of $B_{\rm ex}$ all the way up to $\simeq 2.3$~K and coincides with $B_{\rm ex}$ for higher temperatures. The intensities of the $B=0$ and $B\geq B_{\rm ex}$ peaks are proportional to the volume fractions of the superconducting ($B_{\rm S}=0$) and the normal state ($B_{\rm N}=B_{\rm c}$) domains. The disappearance of the $B=0$ peak above $2.3$~K correspond to the transition of the sample into the normal state [$T_{\rm c}(B=20 \ {\rm mT})\simeq 2.3$~K]. The position of the $B > B_{\rm ex}$ peak represents the temperature evolution of the thermodynamical critical field $B_{\rm c}$ [red dashed line in Fig.~\ref{fig:experiment}~(d)].

Note that our $\mu$SR data exclude the possibility of type-II superconductivity in Bi-II. Additionally, the zero-temperature critical field was found to be half the value of  $B_{\rm c}(0)\simeq 73$~mT reported in Ref.~\onlinecite{Li_PRB_2017}. Field scans at $T=0.25$, 2.1 and 3.0~K with 1~mT steps (from 0.3 to 35 ~mT) and temperature scans at $B_{\rm ex}=3$, 10, 20 and 30 and 35~mT with 0.125~K steps (from 0.25 to 8.0~K) do not show {\it any} FLL-type $\mu$SR response. No superconductivity was detected at $B_{\rm ex}=35$~mT down to the lowest temperature of the experiment ($\simeq0.25$~K) and for all applied fields at $T\geq 4$~K. The fact that no FLL signal was observed above $4.0$~K, suggests also that the admixture of the Bi-III phase ($T_{\rm c}\simeq 7$~K as is detected in ACS experiment, see the Supplemental part, Ref.~\onlinecite{Supplemntal_part}) is minimal in sample volume. Our results imply, therefore, that within the full range of temperatures ($0.25\leq T\leq 8.0$~K) and fields ($0.3\leq B_{\rm ex} \leq 0.35$~mT) studied, the Bi-II phase of elemental Bismuth behaves as a typical {\it type-I superconductor}.

The temperature  dependence of the thermodynamical critical field $B_{\rm c}$, as determined from the measured field value in the normal-state domain [$B_{\rm c}=B_{\rm N}$, see Figs.~\ref{fig:experiment}~(b), (c), and (d)], is shown in Fig.~\ref{fig:experiment}~(e). The points are obtained with several applied fields ($B_{\rm ex}=3$, 10, 20, and 30~mT) and they overlap within certain temperature and field regions. 
The reason for such overlapping is caused by the intermediate state formation condition: $B_{\rm c}(T)\; (1-n)\leq B_{\rm ex} \leq B_{\rm c}(T)$, showing that {\it similar} $B_{\rm c}(T)$  can be obtained for different $B_{\rm ex}$'s.\cite{Gladisch_HypInt_1979, Grebinnik_JETP_1980, Egorov_PhysB_2000, Egorov_PRB_2001}

\begin{figure*}[tb]
\includegraphics[width=1.0\linewidth]{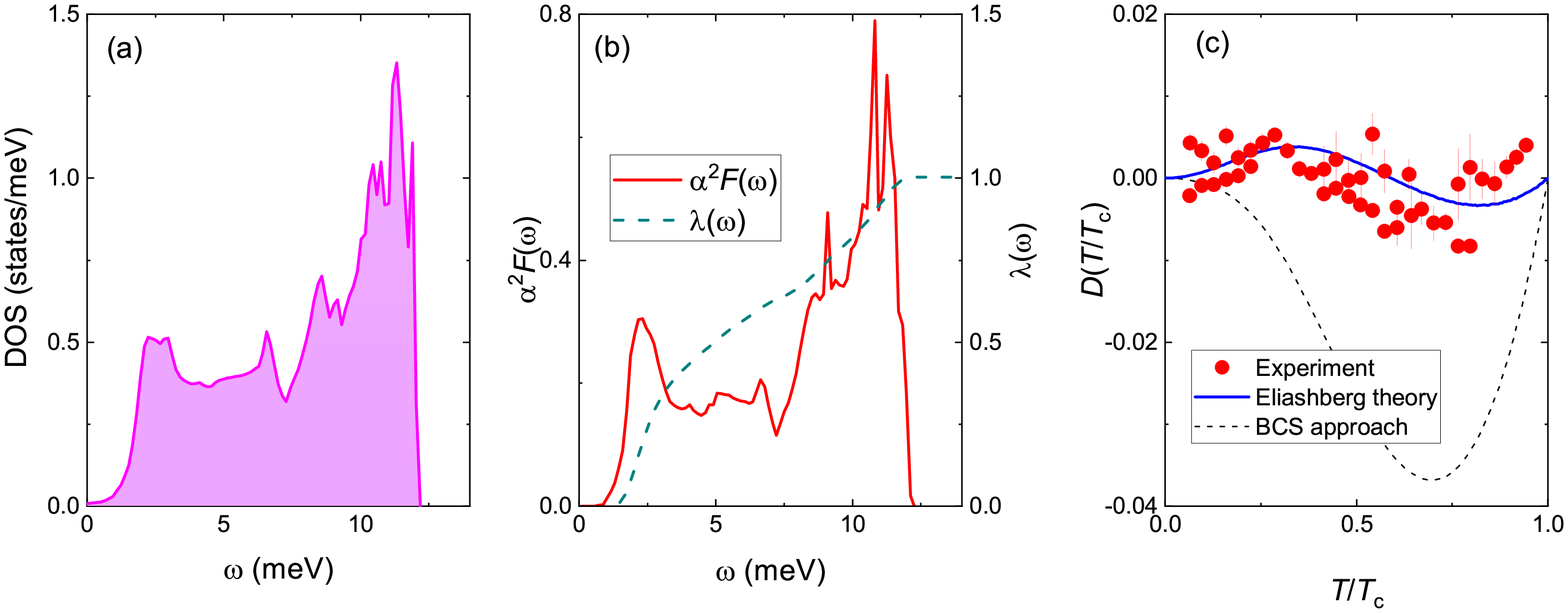}
\vspace{-1.3cm}
\caption{(a) Calculated phonon density of states. (b) Calculated Eliashberg electron-phonon spectral function (red solid line) and integrated electron-phonon coupling strength $\lambda_{\rm el-ph}$ (green dashed line).
(c) The deviation function $D(T/T_{\rm c})=B_{\rm c}(T)/B_{\rm c}(0)-(1-[T/T_{\rm c}]^2)$. The solid blue and dashed line correspond to the Eliashberg and the BCS approach, respectively.
}
 \label{fig:theotc}
\end{figure*}

The obtained experimental data were compared with quantitative predictions based on {\it ab-initio} Eliashberg calculations using Density Functional Theory (DFT). The details of calculations are given in the Supplemental part, Ref.~\onlinecite{Supplemntal_part}. The experimental and calculated material parameters for Bi-II phase are summarized in Table~\ref{tab1}.

Figures~\ref{fig:theotc}~(a) and (b) display the phonon density of states (DOS), the Eliashberg electron-phonon spectral function [$\alpha^2F(\omega)$] and the integrated electron-phonon coupling constant:
$\lambda_{\rm el-ph}(\omega)=2\int_0^\omega \frac{\text{d}\omega'}{\omega'} \alpha^2 F(\omega')$. In the high frequency limit $\lambda_{\rm el-ph}$ was estimated to be $\lambda_{\rm el-ph}(\omega \rightarrow \infty) \simeq 1.02$.
The logarithmically averaged phonon frequency $\omega_{\rm ln}$, representing a characteristic phonon energy mediating the pairing,\cite{al.dy.75} was calculated via:
\begin{equation}
\omega_{\rm ln}=\exp\left(\frac{2}{\lambda_{\rm el-ph}} \int_0^{\infty} \frac{\text{d}\omega}{\omega }\alpha^2F(\omega)\ln\omega\right)
 \label{eq:omega_ln}
\end{equation}
and found to be $\omega_{\rm ln}= 4.69\,$meV.

The dashed line in Fig.~\ref{fig:experiment}~(e) represents the temperature evolution of $B_{\rm c, T}(T)$ computed from the free energy difference between the normal and superconducting states ($\Delta F$) via $B_{\rm c}(T)=\sqrt{-8\pi \Delta F}$ (here after the index 'T' accounts for the parameter obtained from the theory). $\Delta F$ was calculated within the strong-coupling Eliashberg theory following the approach developed by Bardeen and Stephen.\cite{ba.st.64} The transition temperature $T_{\rm c, T}=3.95$~K and the zero temperature value of the thermodynamical field  $B_{\rm c, T}(0)=36.6$~mT are found. Scaling the $B_{\rm c}(T)$ curve further allows direct comparison with the experimental data. The adjusted curve with $T_{\rm c}\simeq3.922$~K and $B_{\rm c}(0)\simeq 31.97$~mT is shown by the solid line in Fig.~\ref{fig:experiment}~(e).

In order to better visualize the difference between the theory and the experiment, the deviation function $D(T/T_{\rm c})=B_{\rm c}(T)/B_{\rm c}(0)-(1-[T/T_{\rm c}]^2)$ is plotted in Fig.~\ref{fig:theotc}~(c).  For comparison, the weak coupling BCS results are also shown.  Obviously, the BCS theory underestimates the experimental $D(T/T_{\rm c})$ and a significant
improvement is obtained using the Eliashberg theory. Although some quantitative discrepancies
remain, the main features are captured.

Many thermodynamic quantities, like the condensation energy or the specific heat jump $\Delta C(T_{\rm c})/\gamma_{\rm N} T_{\rm c}=C_{\rm eS}(T_{\rm c})/\gamma_{\rm N} T_{\rm c}-1$, can be expressed directly by using the derivative of $D(T/T_{\rm c})$
as follows:\cite{Padamsee_JLTP_1973}
\begin{equation}
\frac{ \Delta C(T_{\rm c})}{\gamma_{\rm N} T_{\rm c}}=\frac{B_{\rm c}(0)^2}{2\pi \gamma_{\rm N} T_{\rm c}^2}
\left[ \left.\frac{\partial  D(T/T_{\rm c})}{\partial ([T/T_{\rm c}]^2)} \right|_{(T/T_{\rm c})^2=1}-1 \right]^2 .
\end{equation}
Here $\gamma_{\rm N}$ is the electronic specific heat coefficient in the normal state (see the Supplemental part, Ref.~\onlinecite{Supplemntal_part}, for $\gamma_{\rm N}$ estimate) and $C_{\rm eS}(T)/\gamma_{\rm N} T_{\rm c}$ is the
electronic specific heat in the superconducting state. We proceed with the direct numerical calculation of $C_{\rm eS}(T)/\gamma_{\rm N} T_{\rm c}$
within the Eliashberg theory (see the Supplemental part, Ref.~\onlinecite{Supplemntal_part}). The heat capacity jump ${\Delta C(T_{\rm c})}/{\gamma_{\rm N} T_{\rm c}}\simeq2.40$ was found, which is large in comparison with the universal BCS value of $1.43$. Such a large jump in the specific heat for Bi-II is certainly accessible for calorimetric measurements.

To conclude, the superconductivity in the Bi-II phase of elemental Bismuth was studied experimentally by means of muon-spin rotation, as well as theoretically using the Eliashberg theory in combination with Density Functional Theory calculations. Experiments reveal that the magnetic induction in the cylindrical Bi-II sample is separated into normal state and superconducting  domains thus suggesting that Bi-II is a  superconductor of type-I. The transition temperature and the zero temperature thermodynamic critical field were found to be $T_{\rm c}\simeq 3.92$~K and $B_{\rm c}(0)\simeq31.97$~mT, respectively. The electronic, and the superconducting properties of Bi-II were computed from first principles.  Following the phenomenological approach of Carbotte, Ref.~\onlinecite{Carbotte_RMP_1990}, the strong coupling corrections were embodied via the retardation parameter  $k_B T_{\rm c}/\omega_{\rm ln}$. Including retardation effects, the Eliashberg theory provides better agreement with the experimental data than  the weak coupling BCS approach. The theory values for the critical temperature ($T_{\rm c, T}\simeq 3.95$~K) and  the zero temperature critical field $B_{\rm c, T}(0)=36.6$~mT, as well as the temperature evolution of $B_{\rm c}(T)$ are in agreement with the experiment. The specific heat jump, as estimated from the deviation function $D(T/T_{\rm c})$, was found to be  ${\Delta C(T_{\rm c})}/{\gamma_{\rm N} T_{\rm c}}=2.40$, which is  large in comparison with the universal BCS value of $1.43$. The  {\it ab-initio} calculations result in  the value of the retardation parameter $k_B T_{\rm c}/\omega_{\rm ln} \approx 0.07$  and put Bi-II in the category of intermediate coupling superconductors, being away from the very strong coupling limit
$k_B T_{\rm c}/\omega_{\rm ln} \approx 0.25$. Finally, our analysis reveals that the Cooper pairing in Bi-II is a consequence of balance between the electron-phonon attraction and a significant direct Coulomb repulsion.  Compared to our previous study of Bi-III,\cite{Khasanov_PRB-Bi-III_2018}, the retardation effects in Bi-II were found to be less efficient than in Bi-III. While Bi-III is a type-II strong-coupled  superconductor,\cite{Brown_ScAdv_2018,Brown_Thesis_2017,Khasanov_PRB-Bi-III_2018} the Bi-II and Bi-I are a type-I superconductors with the intermediate  (present study) and weak-coupling (Ref.~\onlinecite{Mata-Pizon_Plos_2016}) strength, respectively (see also the Table~\ref{tab1} summarizing experimental and calculated material parameters for various Bismuth phases). In this respect the high pressure $\mu$SR experiments, as those presented here and in Ref.~\onlinecite{Khasanov_PRB-Bi-III_2018} on elemental Bi, are essential tools to elucidate the nature of the interplay between structural and superconducting phases in conventional superconductors.

This work was performed at the Swiss Muon Source (S$\mu$S), Paul Scherrer Institute (PSI, Switzerland). The work of GS is supported by the Swiss National Science Foundation, grants $200021 \_ 149486$ and $200021 \_ 175935$. M.M.R. acknowledges the support from the Ministry of Education, Science, and Technological Development of the Republic of Serbia under project ON171017. Numerical simulations were run on the PARADOX supercomputing facility at the Scientific Computing Laboratory of the Institute of Physics Belgrade.
L.C. and A.\"O acknowledges DFG for the financial support through TRR80/F6
project.
S.S. acknowledges financial support by the Swedish Research Council and computational facilities provided by the Swedish National Infrastructure for Computing at the National Supercomputer Centers in Link\"oping and Ume\aa.


\begin{thebibliography}{99}
%
\bibitem{Shubnikov_Nature_1930} L. Shubnikov and W.J. de Haas, Nature {\bf 126}, 500 (1930).

\bibitem{deHaas_PNRAS_1930} W.J. de Haas and P.M. van Alphen, Proc. Netherlands Roy. Acad. Sci. {\bf 33}, 1106 (1930).

\bibitem{Yang_PRB_2000} F. Yang, K. Liu, K. Hong, D. Reich, P. Searson, C. Chien, Y. Leprince-Wang, K. Yu-Zhang, and K. Han, Phys. Rev. B {\bf 61}, 6631 (2000).

\bibitem{Hofmann_PiSS_2006} P. Hofmann, Progress in Surface Science {\bf 81}, 191 (2006).

\bibitem{Norman_book_1998} N.C. Norman, {\it Chemistry of Arsenic, Antimony and Bismuth}, Springer Science$+$Business Media (1998).

\bibitem{Degtyareva_HPR_2005} O. Degtyareva, M. I. MCMahon, and R. J. Nelmes, High Pressure Res. {\bf 24}, 319 (2004).

\bibitem{Brandt_JETP_1963} N. B. Brandt and N. I. Ginzburg, Sov. Phys. JETP {\bf 17}, 326 (1963).

\bibitem{Buckel_PL_1965} W. Buckel and J. Wittig, Phys. Lett. {\bf 17}, 187 (1965).

\bibitem{Ilina_JETP_1970} M. A. Il'ina and E. S. Itskevich, Sov. Phys. JETP Lett. {\bf 11}, 218 (1970).

\bibitem{Lotter_EPL_1988} N. Lotter and J. Wittig,  Europhys. Lett. {\bf 6}, 659 (1988).

\bibitem{Du_PRL_2005} X. Du, S. W. Tsai, D. L. Maslov, and A. F. Hebard, Phys. Rev. Lett. {\bf 94}, 166601 (2005).

\bibitem{Prakash_Science_2017} O. Prakash, A. Kumar, A. Thamizhavel, S. Ramakrishnan, Science {\bf 355}, 52 (2017).

\bibitem{Li_PRB_2017} Y. Li, E. Wang, X. Zhu, and H.-H. Wen, Phys. Rev. B {\bf 95} 024510, (2017).

\bibitem{Brown_ScAdv_2018} P. Brown, K. Semeniuk, D. Wang, B. Monserrat, C.J. Pickard, and F.M. Grosche, Science Advances  {\bf 4}, eaao4793 (2018).

\bibitem{Brown_Thesis_2017} P. Brown,  {\it High-pressure states of Bismuth}. University of Cambridge	Dept. of Physics, PhD thesis, (2017).

\bibitem{Khasanov_PRB-Bi-III_2018} R. Khasanov, H. Luetkens, E. Morenzoni, G. Simutis, S. Sch\"{o}necker, A. \"{O}stlin, L. Chioncel, and A. Amato, Phys. Rev. B {\bf 98}, 140504(R) (2018).

\bibitem{Bundy_PR_1958} F. Bundy, Phys. Rev. {\bf 110}, 314, (1958).

\bibitem{Klement_PR_1963} W. Klement, A. Jayaraman, and G.C. Kennedy, Phys. Rev. {\bf 131}, 632 (1963).

\bibitem{Il'ina_PSS_1967} M.A. Il'ina and E.S. Itskevich, Sov. Phys.-Solid State {\bf 8}, 1873 (1967).

\bibitem{Compy_JAP_1970} E.M. Compy, Journal of Applied Physics {\bf 41}, 2014 (1970).

\bibitem{Yomo_JPSJ_1972} S. Yomo, N. Mari, and T. Mitsui, J. Phys. Soc. Jpn. {\bf 32}, 667 (1972).

\bibitem{Homan_JPCS_1975} C.G. Homan, Journal of Physics and Chemistry of Solids {\bf 36}, 1249 (1975).


\bibitem{Sharvin_JETP_1958} I.V. Sharvin, Sov. Phys. JETP {\bf 6}, 1031 (1958).

\bibitem{Dorsey_PRB_1998} A.T. Dorsey and R.E. Goldstein, Phys. Rev. B {\bf 57}, 3058 (1998).


\bibitem{Supplemntal_part} See Supplemental Information  at $<$http://link.aps.org/supplemental/xxxxx$>$ for description of the experimental techniques, the data analysis procedure, and the details of the Density Functional Theory calculations. The Suppelemntal part includes Refs.~\onlinecite{Murata_RSI_1997, Eiling_JPF_1981, Suter_MUSRFIT_2012, Khasanov_PRL_2010, Yaouanc_book_2011, Brandt_LTP_1977, Brandt_LTP_1988, Brandt_PRB_1988, Rammer_PhysC_1991, Brandt_PRB_2003, Laulajainen_PRB_2006, gi.ba.09, ba.gi.01, pi.82, ma.88, ka.75}.

\bibitem{Khasanov_PRB_2011} R. Khasanov, S. Sanna, G. Prando, Z. Shermadini, M. Bendele, A. Amato, P. Carretta, R. De Renzi, J. Karpinski, S. Katrych, H. Luetkens, and N. D. Zhigadlo, Phys. Rev. B {\bf 84}, 100501(R) (2011).

\bibitem{Khasanov_HPR_2016} R.~Khasanov, Z. Guguchia, A.~Maisuradze, D.~Andreica, M.~Elender, A.~Raselli, Z.~Shermadini, T.~Goko, E.~Morenzoni, A.~Amato, High Pressure Res. {\bf 36}, 140 (2016).

\bibitem{Shermadini_HPR_2017} Z. Shermadini,  R. Khasanov, M. Elender, G. Simutis, Z. Guguchia, K.V. Kamenev, A. Amato, High Pressure Res. {\bf 37}, 449 (2017).

\bibitem{Collan_PRL_1969} H.K. Collan, M. Krusius, and G.R. Pickett, Phys. Rev. Lett. {\bf 23}, 11 (1969).

\bibitem{Mata-Pizon_Plos_2016} Z. Mata-Pinz\'{o}n Z, A.A. Valladares, R.M. Valladares, and A. Valladares, PLoS ONE {\bf 11}, e0147645 (2016).

\bibitem{Maisuradze_JPCM_2009} A. Maisuradze, R. Khasanov, A. Shengelaya, and H. Keller, J. Phys.: Condens. Matter {\bf 21}, 075701 (2009).

\bibitem{Khasanov_PRB_2016} R. Khasanov,  H. Zhou, A. Amato, Z. Guguchia, E. Morenzoni, X. Dong, G. Zhang, and Z.-X. Zhao,
Phys. Rev. B {\bf 93}, 224512 (2016).

\bibitem{Tinkham_75} M. Tinkham, {\it Introduction to Superconductivity} (Krieger Publishing company, Malabar, Florida, 1975).

\bibitem{Huebener_Book_1979} R.P. Huebener, {\it Magnetic Flux Structures in Superconductors} (Springer-Verlag, New York, 1979).

\bibitem{Prozorov_PRL_2007} R. Prozorov, Phys. Rev. Lett. {\bf 98}, 257001 (2007).

\bibitem{Prozorov_NatPhys_2008} R. Prozorov, A.F. Fidler, J.R. Hoberg, and P.C. Canfield, Nature Physics {\bf 4}, 327 (2008).

\bibitem{Gladisch_HypInt_1979} M. Gladisch, D. Herlach, H. Metz, H. Orth, G. zu Putlitz, A. Seeger, H. Teichler, W. Wahl, and W. Wigand, Hyperfine Interact. {\bf 6}, 109 (1979).

\bibitem{Grebinnik_JETP_1980} V.G. Grebinnik, I.I. Gurevich, V.A. Zhukov, A.I. Klimov, L.A. Levina, V.N. Maiorov, A.P. Manych, E.V. Mel'nikov, B.A. Nikol'skii, A.V. Pirogov, A.N. Ponomarev, V.S. Roganov, V.I. Selivanov, and V.A. Suetin, Sov. Phys. JETP {\bf 52}, 261 (1980).

\bibitem{Egorov_PhysB_2000} V.S. Egorov, G. Solt, C. Baines, D. Herlach, and U. Zimmermann, Physica B {\bf 289-290}, 393 (2000).

\bibitem{Egorov_PRB_2001} V.S. Egorov, G. Solt, C. Baines, D. Herlach, and U. Zimmermann, Phys. Rev. B {\bf 64}, 024524 (2001).

\bibitem{Singh_Arxiv_2019} D. Singh, A. D. Hillier, and R. P. Singh,  arXiv:1901.06492 (2019), unpublished.

\bibitem{Beare_Arxiv_2019} J. Beare, M. Nugent, M. Wilson, Y. Cai, T. Munsie, A. Amon, A. Leithe-Jasper, Z. Gong, S. Guo, Z. Guguchia, Y. Grin, Y. Uemura, E. Svanidze, and G. Luke,  arXiv:1902.00073 (2019), unpublished.

\bibitem{al.dy.75} P.B. Allen and R.C. Dynes, Phys. Rev. B {\bf 12}, 905 (1975).

\bibitem{ba.st.64} J. Bardeen and M. Stephen, Phys. Rev. {\bf 138}, A1485 (1964).

\bibitem{Padamsee_JLTP_1973} H. Padamsee, J.E. Neighbor, and  C.A. Shiffman, J. Low Temp. Phys. {\bf 12}, 387 (1973).

\bibitem{Carbotte_RMP_1990} J.P. Carbotte, Rev. Mod. Phys. {\bf 62}, 1027 (1990).

\bibitem{Murata_RSI_1997} K. Murata, H. Yoshino, H.O. Yadev, Y. Honda,
and N. Shirakawa, Rev. Sci. Instrum. \textbf{68}, 2490 (1997).

\bibitem{Eiling_JPF_1981} A. Eiling and  J.C. Schilling, J Phys F: Met Phys. {\bf 11}, 623 (1981).

\bibitem{Suter_MUSRFIT_2012} A. Suter and B. M. Wojek, Phys. Procedia {\bf 30}, 69 (2012).

\bibitem{Khasanov_PRL_2010} R. Khasanov, M. Bendele, A. Amato, K. Conder, H. Keller, H.-H. Klauss, H. Luetkens, and E. Pomjakushina, Phys. Rev. Lett. {\bf 104}, 087004 (2010).

\bibitem{Yaouanc_book_2011} A. Yaouanc, and P. Dalmas de R\'{e}otier,
\textit{Muon Spin Rotation, Relaxation and Resonance: Applications
to Condensed Matter} (Oxford University Press, Oxford, 2011).

\bibitem{Brandt_LTP_1977} E.H. Brandt, J. Low Temp. Phys. {\bf 26}, 709 (1977).

\bibitem{Brandt_LTP_1988} E.H. Brandt, J. Low Temp. Phys. {\bf 73}, 355 (1988).

\bibitem{Brandt_PRB_1988} E. H. Brandt, Phys. Rev. B {\bf 37}, 2349(R) (1988).

\bibitem{Rammer_PhysC_1991} J. Rammer, Physica C {\bf 177}, 421 (1991).

\bibitem{Brandt_PRB_2003} E. H. Brandt, Phys. Rev. B {\bf 68}, 054506 (2003).

\bibitem{Laulajainen_PRB_2006} M. Laulajainen, F.D. Callaghan, C.V. Kaiser, and J. Sonier, Phys. Rev. B {\bf 74}, 054511 (2006).

\bibitem{gi.ba.09} P. Giannozzi {\it et al.}, J. Phys.: Condens. Matter  {\bf 21}, 395502 (2009).

\bibitem{ba.gi.01} S. Baroni, S. de Gironcoli, A. Dal Corso, and P. Giannozzi, Rev. Mod. Phys. {\bf 73}, 515 (2001).

\bibitem{pi.82} W.E. Pickett, Phys. Rev. B {\bf 26}, 1186 (1982).

\bibitem{ma.88} F. Marsiglio, M. Schlossmann, and J. P. Carbotte, Phys Rev. B {\bf 37}, 4965 (1988).

\bibitem{ka.75} A.E. Karakozov, E.G. Maksimov, and S.A. Mashkov, Sov. Phys. JETP {\bf 41}, 971 (1976).




\end{thebibliography}
\end{document}